FUNCTIONALIZATION OF REDUCED GRAPHITE OXIDE SHEETS WITH A ZWITTERIONIC SURFACTANT

B. Martín-García, M. Mercedes Velázquez*

*Departamento de Química Física, Facultad de Ciencias Químicas. Universidad de Salamanca, E-37008-Salamanca, Spain*

F. Rossella, V. Bellani

*Dipartimento di Fisica "A. Volta", Università di Pavia, I-27100-Pavia, Italy*

E. Diez

*Laboratorio de Bajas Temperaturas, Facultad de Ciencias Físicas. Universidad de Salamanca, E-37008-Salamanca, Spain*

J. L. G. Fierro

*Instituto de Catálisis y Petroleoquímica, CSIC, E-28049-Cantoblanco, Madrid, Spain*

J. A. Pérez-Hernández

*Centro de Láseres Pulsados Ultraintensos (CLPU), E-37008-Salamanca, Spain*

J. Hernández-Toro

*Servicio Láser, Universidad de Salamanca, E-37008-Salamanca, Spain*

S. Claramunt and A. Cirera

*MIND-IN2UB, Departament d'Electrònica, Universitat de Barcelona, E-08028. Barcelona, Spain.*

| | |
|---|---|
| Corresponding author: | M. Mercedes Velázquez |
| Corresponding address: | Departamento de Química Física<br>Facultad de Ciencias Químicas<br>Universidad de Salamanca<br>Plaza de los Caídos s/n<br>37008 Salamanca, SPAIN |
| Fax: | 00-34-923-294574 |
| E-mail: | mvsal@usal.es |

**Abstract**

Films of a few layers in thickness of reduced graphite oxide (RGO) sheets functionalized by the zwitterionic surfactant N-dodecyl-N,N-dimethyl-3-ammonio-1-propanesulfonate (DDPS) are obtained by using the Langmuir–Blodgett method. The quality of the RGO sheets is checked by analyzing the degrees of reduction and defect repair by means of X-ray photoelectron spectroscopy, atomic force microscopy (AFM), field-emission scanning electron microscopy (SEM), micro-Raman spectroscopy, and electrical conductivity measurements. A modified Hummers method is used to obtain highly oxidized graphite oxide (GO) together with a centrifugation-based method to improve the quality of GO. The GO samples are reduced by hydrazine or vitamin C. Functionalization of RGO with the zwitterionic surfactant improves the degrees of reduction and defect repair of the two reducing agents and significantly increases the electrical conductivity of paperlike films compared with those prepared from unfunctionalized RGO.

**Title running head**: LB films of functionalized graphitic samples.

## 1. Introduction

In the last few years graphene has received enormous interest due to its unique properties, such as high thermal conductivity, Young modulus and intrinsic carrier mobility.[1-6] Owing to these remarkable properties, graphene and its derivatives are promising candidates for the fabrication of electronic devices[7], or reinforced filler in composites with applications in medicine.[8] The success of graphene in technological applications is related to the availability of production methods for the synthesis of large amount of material at low cost. The main challenge in the production of graphene is the reduction of the aggregation by π-π stacking. Currently, there are three main routes to produce high-quality graphene flakes: micromechanical exfoliation[9], chemical vapor deposition of hydrocarbons on metal substrates[10] and thermal reduction of silicon carbide[11]. Micromechanical exfoliation gives samples with the highest charge carrier mobility[12] although the process is time consuming, while the other routes require very high temperatures. The preparation of graphene by chemical methods[5] based on the chemical oxidation of graphite followed by a reduction process, is considered one of the most attractive method to obtain graphene, because it is low-cost and scalable method, versatile for processing or chemical functionalization. However, chemical oxidation disrupts the electronic structure of graphene by introducing O-containing groups such as carbonyl, epoxy and hydroxyl groups in the network, which cannot be completely removed by chemical reduction. On the other hand, the reduced graphite oxide tends to agglomerate, making further processing quite difficult. Currently, these problems can be overcome by the functionalization of graphite oxide (GO) with different stabilizers.[6, 13-16] Several ionic surfactants[14] were used with this purpose. The surfactant molecule not only plays an important role in controlling the restacking of flakes, but also functionalizes the graphene sheets. The functionalized sheets can be used in several applications such as solar energy storage[17] or Li-ion batteries[18]. On the other hand, non-covalent functionalization of graphene provides an alternative approach to modify the material properties allowing a better processing and interactions with other compounds without altering the chemical structure of graphene.[19]

In this work we developed a novel approach to obtain zwitterionic surfactant functionalized graphene using the surfactant dodecyl dimethyl ammonium propane sulphonate (DDPS). The advantage of this approach is that the zwitterionic surfactants exhibit greater adsorption onto hydrophobic surfaces such as graphite than the ionic surfactants [20, 21], therefore, we expect that the functionalization of RGO with DDPS provides better quality of the RGO sheets than that obtained with ionic surfactants. On the other hand, DDPS molecules attached on RGO can bind metal cations[17] and different kind of polymers[22-25] producing nanocomposites with potential applications in the construction of photovoltaic devices[17] and sensors[26], respectively. Moreover, the zwitterionic surfactants present higher tolerance to extremes of pH, strong electrolytes, and oxidizing and reducing agents than the ionic ones,[20] therefore they have utility in the fabrication of devices subjected to extreme conditions.

To evaluate the quality of the reduced sheets we used the criteria proposed by Luo *et al*.[27] These criteria consist of analyzing simultaneously the reduction degree, the defect repair degree and the electric conductivity of the graphitic material by using atomic force microscopy (AFM), X-Ray photoelectron spectroscopy (XPS), UV-vis absorption spectroscopy, Micro-Raman spectroscopy and electric conductivity measurements. We selected hydrazine and Vitamin C as reducer agents because they have demonstrated render high quality RGO sheets[27-30]. With respect to other reducer agents[27-31], hydrazine is one of the most widely used, while Vitamin C is an effective green chemistry alternative to hydrazine [28-30, 32, 33] and renders flakes with less structural defects than hydrazine[32,33].

On the other hand, the implementation of graphene obtained by chemical reduction in device fabrication requires uniform and reproducible deposition methodology. Several techniques such as drop-casting[34] or spin-coating[35] have been used to obtain graphene films onto solid wafers; however, these methodologies often result in non-uniform film thickness on solid substrates[3]. An effective method to transfer water-insoluble molecules[36] or nano-materials[37, 38] from the air-water interface onto solids is the Langmuir-Blodgett (LB) methodology. This method presents several advantages since it allows a great control of the inter-particle distance and, consequently, of the inter-particle interactions.[36] Recently this technique has been successfully used for deposition of GO[39] and multilayered graphene films[40]. Therefore, we selected this methodology to deposit no collapsed sheets of our RGO samples onto silicon. The morphology and quality of the LB deposited sheets were studied by means of AFM, field emission scanning electron microscopy (FE-SEM) and Micro-Raman spectroscopy.

## 2. Results and Discussion

**2.1 Graphite oxide.** The GO was synthesized by a slight modification of the Hummers' method[41], see Experimental Section. In order to evaluate the quality of the starting material, the GO samples were characterized by XPS and the results are reported in Table 1. The $C_{1s}$ core-level spectrum of GO can be fitted by three components centered at 284.8, 286.3 and 288.6 eV, being the first two more intense than the third one (Table 1). These peaks are assigned to C-C bonds in aromatic networks, to C-O bonds in alcohols or epoxy groups and to $COO^-$ structures, respectively.[42] In addition, the $O_{1s}$ core-level spectrum shows two peaks at 531.4 and 533.6 eV. The first one is assigned to oxygen in C=O groups while the later comes from C-O bonds[42] and it is more intense than the first one. This observation indicates that the epoxy and hydroxyl groups are the major components, in agreement with results obtained elsewhere[42]. The O/C ratio value calculated from XPS was 0.62. This value is higher than the values found in literature (from 0.37 to 0.53)[13, 28-30, 43], because we have employed a modification of the Hummers' method to obtain more oxidized GO.

**2.2 Reduced graphite oxide sheets deposited by the Langmuir-Blodgett methodology.** Firstly, we checked the efficiency of the reduction process by analyzing the position of the maximum of the UV-vis absorption spectra of the different samples. As shown in Table 2, the position of the maximum for GO is centered at 230 nm, in agreement with literature.[44] After reduction, the absorption peak red shifts to 264 nm and 265 nm for reduction with hydrazine and Vitamin C, respectively, indicating that both are good reducer agents of GO and that reduction achieved is similar for the two reducer agents.[44]

X-Ray photoelectron spectroscopy (XPS) was used to analyze the reduction degree of the RGO samples. Results obtained for the different samples are summarized in Table 1. Each reported value is an average over at least three spectra, and the standard deviation of these measurements was considered the experimental error.

Figure 1a shows the $C_{1s}$ core-level spectra of different samples. The asymmetric peak of the $C_{1s}$ was fitted to three or four components[13] centered at 284.8 eV (aromatic $sp^2$), 286.9 eV (C-O or C-N), 287.7 eV (C=O) and 290 eV (O-C=O), respectively. From the data reported in Table 1 and shown in Fig. 1a, it can be seen that the most significant change after the GO reduction is the increase in the $sp^2$ component, indicating an increase of the proportion of reduced carbons. Our data also indicate similar $sp^2$ component in samples reduced by hydrazine[27,45,46] and Vitamin C, in good agreement with literature[28], while the O/C atomic ratio of sample reduced by Vitamin C is higher than in RGO reduced by hydrazine (Table 1).

The increase of the O/C ratio is due to the formation of hydrogen bonds between the oxidized product of Vitamin C and the residual oxygen on the RGO.[29] We also recorded the $N_{1s}$ core-level spectra of the different chemically reduced graphite samples (Fig. 1b). The $N_{1s}$ spectrum for RGO reduced with hydrazine presents one peak centered at 400.0 eV while no $N_{1s}$ peak was detected for the RGO reduced by Vitamin C. Other authors demonstrated that the reduction of the epoxy groups using hydrazine is blocked by the formation of the hydrazino alcohol[47], so that the XPS peak at 400.0 eV was ascribed to the nitrogen atoms (amine nitrogen) attached to the RGO sheets[48].

The RGO sheets thus obtained were deposited onto silicon by using the LB methodology. The fabrication of RGO sheets by using the LB methodology requires a detailed characterization of the RGO Langmuir monolayers, which are precursors of the LB films. To this aim we studied the stability of the spreading solutions employed to prepare the Langmuir monolayers and the different states of the RGO Langmuir monolayers, by recording the surface-pressure isotherms (see Supporting Information). Langmuir monolayers corresponding to different surface states were transferred onto silicon by LB. We noticed that the LB films built with the densest Langmuir monolayers ($\geq$ 4 mN $m^{-1}$) consist of flakes with several layers and high roughness (see Supporting Information). Therefore, since we were interested in obtaining no overlapped sheets, we prepared the LB films by transferring dilute RGO Langmuir monolayers ($\pi \leq$ 4 mN $m^{-1}$).

The RGO Langmuir monolayers were prepared by spreading a solution (0.1 mg $mL^{-1}$) of RGO dispersed in chloroform. It is established that sonication of RGO prevents the restacking of the graphene flakes, therefore we prepared the spreading solutions using this method, with sonication times between 75 and 150 min. Solutions stable for several weeks were obtained with sonication times ranging from 90 to 150 min. The Brewster angle microscopy (BAM) of the Langmuir monolayers did not show images of the RGO nanoplatelets. This fact is signature of good-quality RGO flakes, because the graphene monolayer is an almost transparent material (pa 2.3%)[49].

Deposition of RGO by the LB methodology was carried out for all samples except RGO obtained by reduction with hydrazine. In this case, it was not possible to prepare Langmuir monolayers of this material because aggregates were observed in the spreading solution few minutes after preparation. However, we deposited RGO reduced by hydrazine using a fresh and transparent spreading solution. Then, we observed visible aggregates after deposition of RGO on the water by BAM, see Supporting Information. These aggregates were transferred onto silicon by the LB methodology and its morphology analyzed by AFM, see Supporting Information. The film profiles obtained from AFM show aggregates of high roughness (>100 nm) (Supporting Information). Therefore, we decided to reduce *in situ* the sheets of GO deposited on silicon (Figure 2) by exposure to hydrazine vapor[39]. Figures 3a and 3c show high magnification AFM and FE-SEM images of representative flakes corresponding to the RGO obtained by *in situ* reduction of GO with hydrazine vapor. The roughness of the flake determined by AFM was ~ 1.5 nm (Fig. 3b), compatible with a 1-3 graphene layers[50]. To analyze the degree of defect repair after the reduction of GO by hydrazine vapor, we have recorded the Micro-Raman spectra of RGO sheets deposited on silicon (Fig. 3d). The spectrum presents three bands, centered at 1587 $cm^{-1}$ (G band), 1330 $cm^{-1}$ (D band) and 1618 $cm^{-1}$ (D' band), respectively. The D and D' bands are identified as disorder bands and are originated from different mechanisms, intervalley (D) and intravalley (D') resonant Raman scattering.[51, 52] The intensity of the D band depends on the degree and nature of the basal plane disorder because this band requires defects for its activation[53, 54], while its position depends on the incident laser energy[53]. The D band is observed in chemical derived flakes because oxidation and reduction processes seriously alter the basal plane of graphene. As the $I_D/I_G$ ratio is a measure of the defects on the $sp^2$ bonding character[54], we can note from data in Table 2 that the $I_D/I_G$ ratio value is high, indicating the presence of non-reduced groups or disorder introduced by the functionalization of the graphene sheets by the nitrogen compounds.

Figures 4a, 4b and 4c show the FE-SEM and AFM images of different LB deposited sheets obtained by transferring the monolayers of RGO (1 mN $m^{-1}$) produced by the reduction with Vitamin C. We calculated the average roughness value of the sheets from AFM. The value found, ~ 2 nm, is similar to that obtained for RGO reduced by hydrazine. Therefore, we use the SEM images to evaluate the morphology of the flakes and AFM to determine its roughness. Comparison between the SEM images of RGO flakes obtianyde by reduction with hydrazine and Vitamin C show that Vitamin C provides larger flakes than hydrazine. This could be due to differences between the mechanisms involved in the reduction processes. Thus, cracking of the carbon network in the reduction process promoted by hydrazine may be rationalized taking into account the thermal instability of aziridines[43] leading to carbon nitrogen 1,3-dipoles[56] which may be hydrolyzed under reaction conditions giving rise to complex mixtures of substituted hydrazine and carbonyl compounds[57].

The Raman spectrum of flakes reduced with Vitamin C also exhibits the G and D bands (Fig. 4d). However, the $I_D/I_G$ ratio is smaller than in the case of RGO samples obtained by reduction with hydrazine (see Table 2). This fact indicates an increase in the size of the $Csp^2$ domains for RGO samples reduced with Vitamin C. However, at the same time, the XPS results indicate a similar amount of $Csp^2$ domains for the two samples. This suggests that even though the number of $Csp^2$ domains is similar, the reduction with Vitamin C provides larger $Csp^2$ domains than the reduction with hydrazine[58, 59]. The $I_D/I_G$ ratio for RGO reduced by Vitamin C found in this work is smaller (0.67) than those reported in literature (~1)[29, 30], suggesting that the size of the $sp^2$ domains is larger in our sample. Taking into account the high oxidation degree of our GO samples with respect to other works[29, 30], our results suggest that a higher oxidation degree of starting GO renders larger $sp^2$ domains when the reduction is carried out by Vitamin C, while no significant changes are observed when hydrazine is employed as reducer agent.

**2.3 Langmuir-Blodgett deposited sheets of reduced graphite oxide functionalized with the zwitterionic surfactant DDPS.** To obtain functionalized RGO, we have carried out the reduction with hydrazine and Vitamin C in the presence of the DDPS surfactant (see Experimental Section for details). We investigated the efficiency of the reduction process assisted by DDPS surfactant analyzing both, the UV-vis absorption spectrum and the $C_{1s}$ core-level spectra, as reported in Table 2 and Fig. 1a, respectively. As can be seen, the red shift of the peak positions of the UV-vis absorption spectra increases for samples reduced in surfactant solutions indicating better reduction efficiency. This observation is supported by the XPS results, that show a higher percentage of $Csp^2$ for RGO samples reduced in the presence of surfactant, respect to the percentage obtained in the absence of the surfactant (Tables 1 and 2). Therefore, we conclude that the surfactant improves the reduction of the GO.

To corroborate the functionalization of the RGO flakes by the surfactant DDPS, we analyzed the XPS spectra of the different samples. As can be seen in the XPS spectra, the RGO samples prepared in surfactant solutions present the $N_{1s}$ and $S_{2p}$ core-level spectra at binding energies of 402.0 eV (Fig. 1b) and 168.0 eV, respectively (Table 1). In order to assign these peaks we compare these results with the XPS spectrum of the surfactant DDPS, observing that the $N_{1s}$ core-level spectrum of the DDPS surfactant (Fig. 1b) shows a symmetric peak centered at a binding energy of 402.3 eV which is similar to that recorded for RGO samples reduced in the presence of the surfactant DDPS (see Fig. 1b and Table 1). The $S_{2p}$ core-level spectrum of the surfactant also presents the sulphonate group at the binding energy of 167.5 eV.[13] The small shift in the binding energy of the $S_{2p}$ peak in RGO samples intercalated with DDPS reflects environment effects due to the carbon network. From these results, we can conclude that the synthesis proposed in this work allows obtaining RGO flakes with surfactant molecules attached to the graphitic network.

It is also interesting to note that when the reduction is carried out with hydrazine assisted by the surfactant DDPS, the $N_{1s}$ peak is asymmetric and can be decomposed into two peaks centered at 399.8 eV and 402.0 eV, respectively. As commented above, the peak centered at 402.0 eV corresponds to the nitrogen atoms of the surfactant attached to the RGO sheets, while the peak at 399.8 eV to the nitrogen atoms attached to the RGO reduced with hydrazine[48]. Consequently, from our results, it is possible to conclude that although the surfactant cannot completely inhibit the defects introduced by nitrogen groups incorporated to the network by the reduction using hydrazine, improves the quality of the RGO flakes. Thus, the N/C atomic ratio decreases from 0.039 for hydrazine to 0.004 for hydrazine dissolved in DDPS.

Figure 5a and 5b show the AFM and FE-SEM images of DDPS-functionalized RGO sheets obtained by reduction of GO with hydrazine dissolved in surfactant solutions. The surface pressure of the Langmuir monolayer precursor of the LB film was 1 mN $m^{-1}$. The SEM images show significant differences between the morphology of RGO sheets reduced with hydrazine vapor and with hydrazine dissolved in surfactant solution. Thus, chained nanoplatelets can be observed in the latter. The cross-sectional view of the AFM images indicates that the roughness values of sheets are of 4 nm. This roughness is 2.5 nm higher than the value found for the RGO sheets reduced by hydrazine vapor. However, it is reasonable to assume that the surfactant attached on the network resulted in a thicker film.[60] On the other hand, the formation of the chained sheets suggests lateral attractive interactions between flakes, likely induced by the surfactant molecules attached to the sheets.

Comparison between the AFM and SEM images show some differences. Thus, the AFM images show different lateral shape than the ones obtained by SEM. Taking into account that the AFM images are always the deconvolution of the probe geometry and the shape of the features some image artifacts can be often observed. In our systems these artifacts seem to be more visible in RGO samples obtained by reduction with hydrazine and Vitamin C in the presence of the surfactant DDPS, Figs.5a and 6a, respectively. Attractive interactions between sheets induced by the surfactant push the sheets close enough to obtain AFM images without artifacts. In these situations, the SEM measurements are recommended to obtain information about the real lateral dimensions and density of domains while AFM must be used to determine its roughness [55].

We also analyze the effect of the surfactant DDPS on the degree of defect repair after reduction by means of Micro-Raman spectroscopy. The Raman spectra of RGO functionalized with DDPS and reduced by hydrazine also presents the D and G bands (Fig. 5d). Comparing this spectrum with that of RGO reduced with hydrazine vapor (Fig. 3d), we observe that the $I_D/I_G$ ratio (Table 2) is smaller in the former. This fact, along with the higher percentage of $Csp^2$ detected by XPS for samples functionalized with the surfactant, indicate that the surfactant DDPS improves not only the reduction efficiency but also the degree of defect repair, consistently with results obtained with ionic surfactants[43, 61, 62].

Figure 6a and 6b show the AFM and FE-SEM images of LB films of RGO reduced with Vitamin C functionalized with the surfactant DDPS. As it was explained previously, we use SEM images to obtain information of the flakes morphology and AFM to calculate the roughness of the RGO flakes. SEM images show that, as observed for functionalized RGO samples reduced by hydrazine, interconnected nanoplatelets can be observed. The roughness values of these flakes estimated from AFM are between 3 and 4 nm (Fig. 6c). This value is consistent with the presence of surfactant molecules attached to the carbon network[60] as in the case of functionalized RGO flakes reduced with hydrazine.

Figure 6d shows a representative Raman spectrum of the DDPS-functionalized RGO reduced by Vitamin C, which also presents the D and G bands. From results in Table 2, one can see that the $I_D/I_G$ ratio is slightly higher than that obtained by reducing GO with Vitamin C in the absence of surfactant. In contrast, the percentage of $Csp^2$ is higher when the RGO is functionalized by DDPS. Our results indicate that the surfactant DDPS improves the reduction of GO by Vitamin C increasing the percentage of $Csp^2$, while decreases the size of the $sp^2$ domains as compared with the RGO reduced by Vitamin C.

**2.4 Electric conductivity measurements.** The criterion that reflects both the reduction and the defect repair degrees more directly is the electric conductivity. We measured the electric conductivity by a four-probe setup on the RGO paper-like films[63], and the results obtained for different samples are collected in Table 2. This method was proposed by other authors and has been successfully used to carry out the comparative analysis of the quality of graphene samples.[63]

The conductivity value of RGO sheets reduced by hydrazine agrees very well with values obtained previously by other authors[27, 43]. The values recorded for sheets reduced by Vitamin C can not be compared with values in the literature, because the values found correspond to samples reduced by Vitamin C with a subsequent thermal annealing step that can remove the oxidized form of the Vitamin C attached to the network, increasing the conductivity of sheets[43]. The conductivity values reported in Table 2 show that the RGO sample reduced by Vitamin C is less conductive than the one reduced by hydrazine. Taking into account that the $Csp^2$ percentage are quite similar and the $I_D/I_G$ ratio is smaller for the RGO reduced by Vitamin C, one expects higher conductivity for the sample reduced by Vitamin C. However, the behavior found is the opposite one. This can be tentatively explained if one considers that the residual O-containing groups of the oxidized Vitamin C destroy the graphene structure making the sample less conductive[27]. According to it, the low conductivity value of RGO reduced by Vitamin C can be due to the residual groups containing oxygen attached to the network (evaluated by the O/C ratio).

It is interesting to note that the functionalization with the zwitterionic surfactant DDPS increases the conductivity for samples reduced by both, hydrazine and Vitamin C. This behavior is consistent with the high values found for the percentage of $Csp^2$ (Table 2) and with the decrease of the N/C (hydrazine) and O/C (Vitamin C) ratios (Table 1) for the functionalized RGO. Surfactant stabilization of the GO can be responsible of this behavior. Thus, when the GO flakes are dispersed in the surfactant aqueous solution by the application of ultrasound, the surfactants molecules are adsorbed on the sheets stabilizing the GO exfoliated by sonication. The surfactant adsorbed at the surface avoids the restacking and provides an effective reduction of the GO.

Comparing the electrical conductivity values obtained for RGO functionalized with DDPS surfactant with the values found in the literature for paper-like films of RGO with ionic surfactants, the conductivity of our samples is generally higher than these values[46, 61, 64]. This behavior can be interpreted if one takes into account that the zwitterionic surfactants present higher adsorption on the hydrophobic surfaces than the ionic ones[20, 21], providing better stabilization against the restacking process than that obtained by the ionic surfactants. Besides, the DDPS surfactant improves the conductivity value of RGO paper-like films functionalized with the zwitterionic surfactant CHAPS[64]. Structural differences between these surfactants can be responsible of this behavior. Thus, our results suggest stronger interactions between the hydrophobic moiety of DDPS and RGO than between the hydrophobic part of CHAPS and the RGO carbon network. This fact can be due to the more hydrophilic character of the surfactant CHAPS.[24]

## 3. Conclusions

In this work, we studied the effect of a zwitterionic surfactant in the chemical reduction of GO and self-assembly of RGO sheets. Our results demonstrated that the functionalization with the zwitterionic surfactant DDPS improves the reduction efficiency of hydrazine and Vitamin C measured by the percentage of $Csp^2$; reduces the residual groups attached to the carbon network introduced by the reducer agents in both cases and increases the conductivity of the RGO. Moreover, the electrical conductivity observed in our functionalized samples is generally higher than other surfactant-functionalized RGO paper-like films. Accordingly, we propose the DDPS functionalization of RGO as a good non-covalent functionalization, which increases the electrical conductivity without altering the carbon network. Besides, the functionalization with a zwitterionic surfactant offers a residual charge that can bind the RGO sheets with other materials in order to prepare nanocomposites with multiple applications[65].

## 4. Experimental section

**Materials.** The following materials (Sigma-Aldrich) were used without further purification: $NaNO_3$ (99%), $H_2SO_4$ (98%w), $KMnO_4$ (> 99%), $H_2O_2$ (30%w), hydrazine hydrate (80%w) and Vitamin C (L-ascorbic acid, puriss.). The zwitterionic surfactant, dodecyl dimethyl ammonium propane sulphonate (DDPS) was purified by recrystallization in isopropanol until obtaining constancy in the surface tension value of a surfactant solution of concentration close to the CMC, critical micelle concentration.[66] Millipore Ultra pure water prepared using a combination of RiOs and Milli-Q systems from Millipore was used to prepare the subphase. The solid substrate is As-doped silicon wafers (100) with 300 nm of dry thermal $SiO_2$ thin film in order to enhance the optical contrast of the flakes under white-light illumination[67] and was supplied by Graphene Industries, U.K.

**Synthesis and purification of Graphite Oxide.** The GO was prepared by using a slight modification of the Hummers' oxidation method[41] from natural graphite flakes (Qingdao super graphite, LTD. 99.02 fixed C). In order to obtain high quality GO we used a centrifugation-based purification procedure published elsewhere[39]. In our procedure, to achieve a more oxidized material (GO), in the first oxidation step, the reactive solution ($NaNO_3$/ $H_2SO_4$/ $KMnO_4$/graphite) was stirred at 35ºC overnight until a thick paste was formed. This is the main difference respect to the Hummers' method which spends 30 min in the oxidation process.

In the purification procedure[39], the GO filter cake was dispersed in water by mechanical agitation, and centrifuged at 1000 rpm (90*g*) for 3-5 times, allowing to remove the biggest (visible) particles. The supernatant was treated with high-speed centrifugation at 5000 rpm (3750*g*) for 15 min, in order to eliminate the small GO particles and the water-soluble products of the oxidation process. Finally, the supernatant was centrifuged at 10000 rpm (8163*g*) for 15 min to separate the GO sediment, which was then dried[41] at 40 ºC over phosphorus pentoxide in vacuum.

**Chemical reduction of GO with hydrazine without surfactant.** The GO aqueous dispersion obtained by sonication (1h), was reduced by the Stankovich's method[43] with hydrazine at 100ºC during 24 h. The reduced product was filtered and washed with water and methanol. For comparative purposes, the GO reduction was also carried out by exposure of GO LB deposited sheets to hydrazine vapor (0.1 mL) during 18 h at room temperature.[39] To deposit the GO sheets by LB method, we prepared the GO Langmuir monolayer by spreading the GO dispersion in MeOH:$H_2O$ (5:1 v/v, sonication 30 min) on the air-water interface.[39] The GO sheets were transferred by symmetric barrier compression (50 mm $min^{-1}$) with the substrate into the trough by vertically dipping it up at 2 mm $min^{-1}$ and then it was reduced by hydrazine vapor and later rinsed with water and dried in an oven for 1h (80ºC). The GO LB deposited onto silicon was characterized by FE-SEM and AFM (Figure 2). The AFM image of a GO film shows a roughness ~ 1.7 nm compatible with a bilayer of graphite oxide (Figure 2b).

**Chemical reduction of GO with Vitamin C without surfactant.** To reduce the GO with Vitamin C, the pH of the GO dispersions (0.1 mg $mL^{-1}$, sonication time 1h) was adjusted to 9-10 with an ammonia solution of 25% and then, the reducer agent, Vitamin C (2 mM) was added. The reactive medium was heated at 95ºC during 15 min.[28] The reduced dispersion was processed by filtration with a PVDF membrane (0.2 μm pore size, Filter-lab).

**Chemical reduction of GO assisted by the surfactant DDPS.** The chemical reduction was carried out in the presence of the DDPS surfactant using GO dispersed in surfactant aqueous solutions by stirring (2h) and subsequent sonication (1h) and following reduction procedures previously described. In order to optimize the reduction of GO, the DDPS concentration was modified above and below the CMC (1.1 mg $mL^{-1}$)[66]. Surfactant concentrations below the CMC did not give stable dispersions, while concentrations above 2 mg mL$^{-1}$ after a few days gave rise to small precipitates. Therefore, we keep the surfactant concentration in 1.7 mg mL$^{-1}$ which is close to the CMC of DDPS and form stable and clear GO dispersions during a few weeks. This result indicates that the optimum surfactant concentration value required to stabilize the GO dispersion is close to the CMC. This fact is in good agreement with results obtained for ionic surfactants[61].

After reduction, the RGO was filtered and washed with water and methanol, in order to remove traces of free molecules of surfactant, until the disappearance of foam in the washing solvent; finally the RGO was dried and stored in vacuum.

**Experimental methods.** The pressure-area isotherms of RGO were recorded on a Langmuir Mini-trough (KSV, Finland) and for the Langmuir-Blodgett deposition a KSV2000 System 2 from KSV was used; both were placed on an anti-vibration table. The RGO sheets were transferred from the air-water interface onto silicon by symmetric barrier compression (50 mm min$^{-1}$) with the substrate into the trough by vertically dipping it up at 5 mm min$^{-1}$.

The spreading solution (0.1 mg mL$^{-1}$) was deposited onto the water subphase with a micrometer Hamilton syringe. The syringe precision is 1 μL. The surface pressure was measured with a Pt-Wilhelmy plate connected to an electrobalance. The subphase temperature was maintained at 20.0 ± 0.1ºC by flowing thermostatized water through jackets at the bottom of the trough. The temperature close to the surface was measured with a calibrated sensor from KSV while the water temperature was controlled by means of a thermostat/cryostat Lauda Ecoline RE-106.

AFM images of the LB sheets were obtained in constant repulsive force mode by AFM (Nanotec Dulcinea, Spain) using the WSXM 5.0 program[68]. The microscope was equipped with a rectangular microfabricated silicon nitride cantilever 100 μm height with spring constant of 0.73 mN $m^{-1}$ (Olympus OMCL-RC800PSA) and a Si pyramidal tip (tip radius < 20 nm). The measurements were carried out at ambient conditions using scanning frequency from 0.5 to 1.2 Hz per line. The silicon wafers were cleaned[69] by mild sonication with acetone (30 s), methanol (30 s) and MilliQ® water (30 s) and finally dried in an oven at 90ºC.

Raman scattering measurements were carried out at room temperature with a micro-Raman spectrometer (Horiba Jobin-Yvon Labram RH) with a 100x objective (laser spot ~ 1$\mu m^2$), a spectral resolution ~ 2 $cm^{-1}$ and laser excitation wavelength at 632.81 nm. Accurate calibration was carried out by checking the Rayleigh band and Si band at 0 and 520.7 $cm^{-1}$, respectively. The sample area was scanned with a spatial resolution of approximately 0.5 μm, an acquisition time of a few min at each point, while the laser excitation power was kept below 1 mW to avoid heating[70] .

X-ray photoelectron spectra of powder samples were measured in a VG Escalab 200 R spectrometer (Fisons Instruments, USA) equipped with an excitation source of MgKα (hν = 1253.6 eV) radiation and a hemispherical electron analyzer. High resolution spectra were recorded working at 20 eV analyzer pass energy. The residual pressure in the analysis chamber was maintained under $2 \cdot 10^{-7}$ Pa during the data acquisition[71].

FE-SEM images were acquired with a Nova NanoSEM 230-FEI microscope using the high-resolution detector in high vacuum (HV) mode, usually applying an accelerating voltage of 10 kV.

For electric conductivity measurements we applied a film four-point probe setup[63] with gold electrodes over RGO paper-like films fabricated by filtration using a PVDF membrane filter (0.2 μm pore size, Filter-lab), and washing with MilliQ® water and methanol in order to remove the excess of reducer agents and free-surfactant[63, 72]. Three measurements were carried out for each sample to obtain an average value for the electric conductivity performing by means of a Keithley 4200 SCS Semiconductor Parameter Analyzer. The films thickness was measured in a workstation combining Scanning Electron Microscopy (SEM) and Focused Ion Beam (FIB) technique sample preparation[73] (see Supporting Information). The conductivity error was estimated from the error value of the thickness film measured by SEM.

UV-vis measurements were recorded on a Shimadzu UV-2401 PC spectrometer at room temperature.

## 5. Acknowledgments

The authors thank financial support from ERDF and MEC (MAT 2010-19727, MAT 2007-62666, FIS2009-07880), Junta de Castilla y León (SA049A10-2, SA138A08) and the Foundation Cariplo (project *Quantdev*). A.C. acknowledges financial support from ICREA Academia program. B.M.G. wishes to thank the European Social Fund and Consejería de Educación de la Junta de Castilla y León for her FPI grant. We thank to Ultra-Intense Lasers Pulsed Center of Salamanca (CLPU) for the AFM facility; to Drs. González and Íñiguez-de-la-Torre (Department of Applied Physics, University of Salamanca) for the electric conductivity facility and to Dr. Benito Rodriguez and International Iberian Nanotechnology Laboratory (INL) for the FIB/SEM facility. Dr. F. Bermejo (Department of Organic Chemistry, University of Salamanca) is also acknowledged by helpful discussions of the RGO synthesis.

## Legends

**Figure 1.** (a) The $C_{1s}$ core-level spectra of graphite oxide (GO), and RGO samples reduced with: hydrazine; hydrazine dissolved in DDPS; Vitamin C and Vitamin C dissolved in DDPS solutions. (b) The $N_{1s}$ core-level spectra of dodecyl dimethyl ammonium propane sulphonate (DDPS) and RGO samples reduced with: Vitamin C dissolved in DDPS solutions, hydrazine dissolved in DDPS solutions and hydrazine.

**Figure 2.** FE-SEM (a) and AFM (b) images of a Langmuir monolayer of graphite oxide at 3mN $m^{-1}$ transferred onto silicon by the Langmuir-Blodgett methodology. The corresponding cross section profile is also shown.

**Figure 3.** AFM (a) and FE-SEM (c) images of the Langmuir-Blodgett film of RGO reduced *in situ* with hydrazine. (b) Profile obtained from AFM. (d) Micro-Raman spectrum of the RGO reduced *in situ* by hydrazine.

**Figure 4.** FE-SEM (a, b) and AFM (c) images of the Langmuir-Blodgett film of RGO reduced with Vitamin C. Below the AFM image is the profile obtained from AFM. (d) Micro-Raman spectrum of a representative RGO flake.

**Figure 5.** AFM (a) and FE-SEM (b) images of the Langmuir-Blodgett film of RGO reduced by hydrazine dissolved in DDPS surfactant solution. (c) Profile obtained from AFM and (d) Micro-Raman spectrum of a representative RGO flake.

**Figure 6.** AFM (a) and FE-SEM (b) images of the Langmuir-Blodgett film of RGO reduced by Vitamin C dissolved in DDPS surfactant solution. (c) Profile obtained from AFM and (d) Micro-Raman spectrum of a representative RGO flake.

# Tables

**Table 1.** Binding energies (eV) and O/C and N/C surface atomic ratios of graphitic samples.

| Sample | $C_{1s}$ | $S_{2p}$ | $N_{1s}$ | O/C | N/C |
|---|---|---|---|---|---|
| **GO** | 284.8 (49 ± 2)<br>286.9 (45 ± 2)<br>288.6 (6 ± 0.5) | - | - | 0.618 | - |
| **RGO Hydrazine** | 284.8 (63 ± 2)<br>286.3 (22 ± 2)<br>287.7 (8 ± 1)<br>289.3 (7 ± 0.5) | - | 400.0 | 0.119 | 0.039 |
| **RGO Hydrazine (DDPS)** | 284.8 (82 ± 1)<br>286.3 (16 ± 0.5)<br>289.0 (2 ± 0.5) | 167.8 | 399.8 (60 ± 1)<br>402.0 (40 ± 1) | 0.190 | 0.004 |
| **RGO Vitamin C** | 284.8 (67 ± 1)<br>286.4 (29 ± 0.4 )<br>288.9 (4 ± 0.2) | - | - | 0.339 | - |
| **RGO Vitamin C (DDPS)** | 284.8 (77 ± 6)<br>286.2 (20 ± 3)<br>289.0 (3 ± 0.5) | 168.6 | 402.2 | 0.209 | 0.006 |
| **DDPS** | 284.8 (82)<br>286.2 (18) | 167.5 | 402.3 | 0.183 | 0.056 |

**Table 2.** UV-vis absorption peak position for aqueous GO solutions and different RGO samples dissolved in chloroform. The concentration was kept constant in 0.1 mg $mL^{-1}$. Percentage of C ($sp^2$) and the intensity ratio values found for different graphitic samples. Electric conductivity values of the different RGO paper-like films.

| Sample | $\lambda_{max}$ / nm | % C ($sp^2$) | $I_D/I_G$ | Conductivity / S $m^{-1}$ |
|---|---|---|---|---|
| **Graphite Oxide** | 230 ± 2 | 49 ± 1 | | Insulator |
| **RGO Hydrazine** | 264 ± 2 | 63 ± 2 | --- | 241 ± 29 |
| **RGO Hydrazine vapor** | --- | --- | 1.99 | --- |
| **RGO Hydrazine (DDPS)** | 268 ± 2 | 82 ± 1 | 1.45 | 484 ± 58 |
| **RGO Vitamin C** | 265 ± 2 | 67 ± 1 | 0.67 | 37 ± 4 |
| **RGO Vitamin C (DDPS)** | 266 ± 2 | 77 ± 6 | 0.92 | 580 ± 70 |

## Illustrations

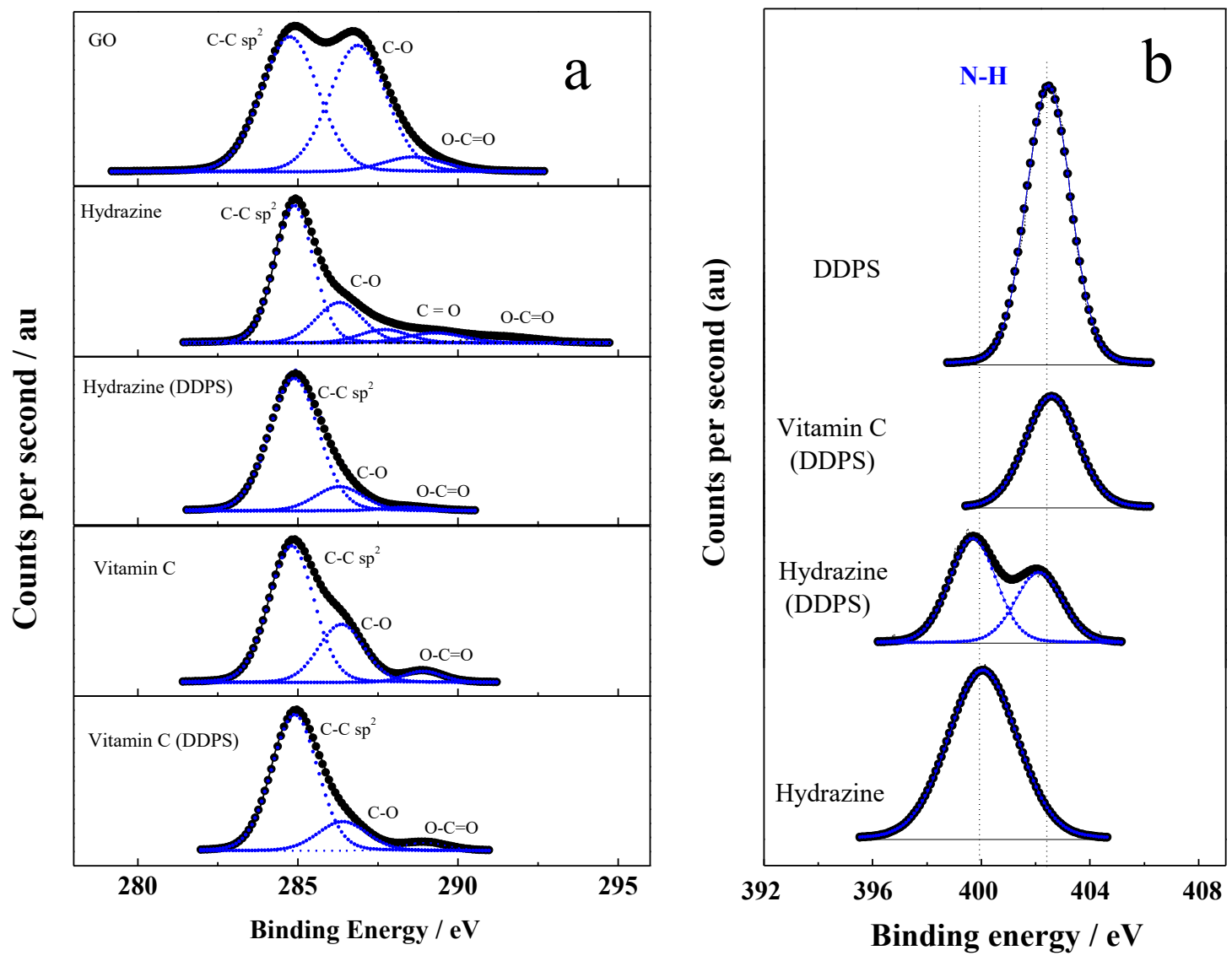

**Figure 1.** (a) The $C_{1s}$ core-level spectra of graphite oxide (GO), and RGO samples reduced with: hydrazine; hydrazine dissolved in DDPS; Vitamin C and Vitamin C dissolved in DDPS solutions. (b) The $N_{1s}$ core-level spectra of dodecyl dimethyl ammonium propane sulphonate (DDPS) and RGO samples reduced with: Vitamin C dissolved in DDPS solutions, hydrazine dissolved in DDPS solutions and hydrazine.

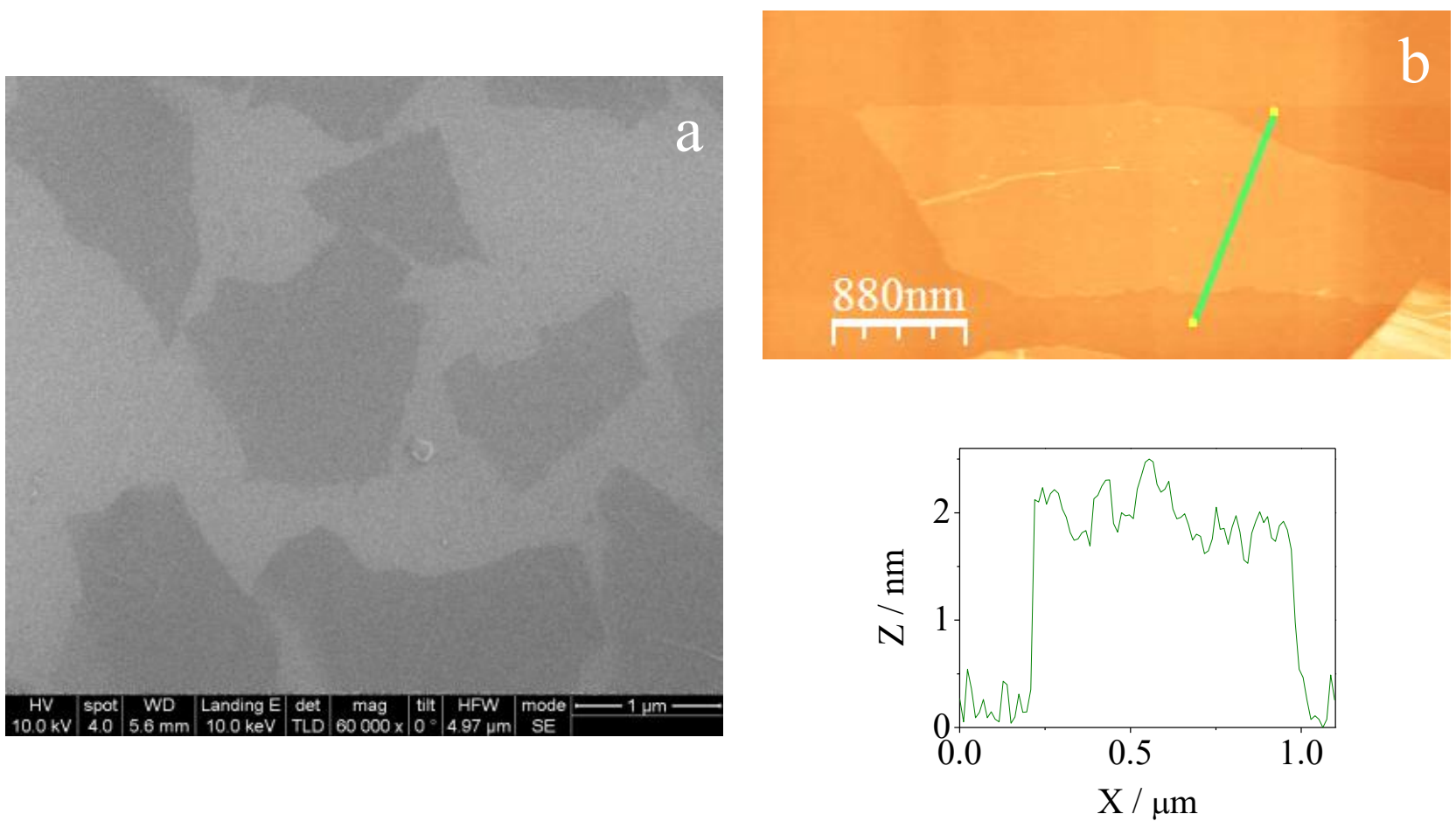

**Figure 2.** FE-SEM (a) and AFM (b) images of a Langmuir monolayer of graphite oxide at 3 mN $m^{-1}$ transferred onto silicon by the Langmuir-Blodgett methodology. The corresponding cross section profile is also shown.

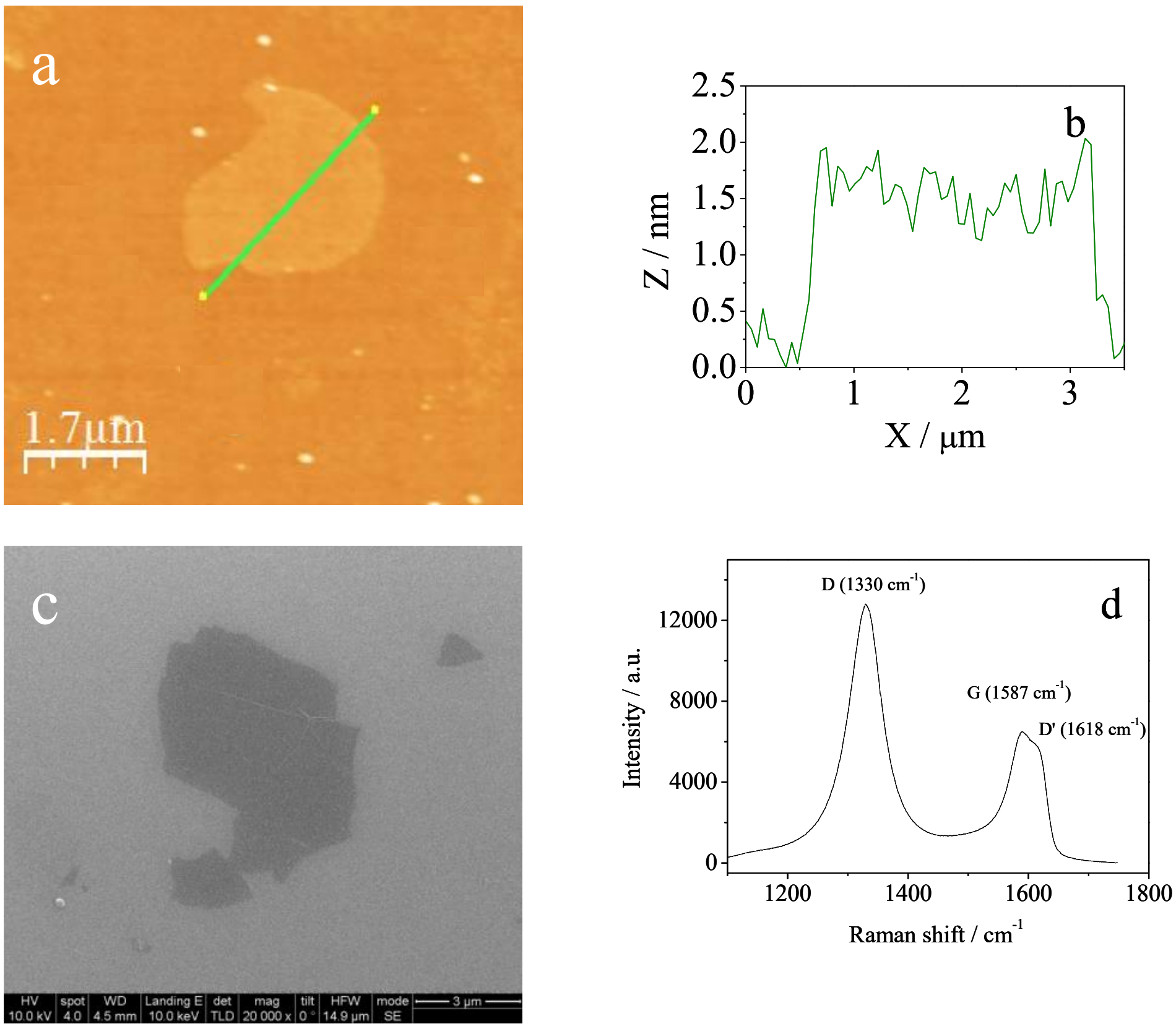

**Figure 3.** AFM (a) and FE-SEM (c) images of the Langmuir-Blodgett films of RGO reduced *in situ* with hydrazine. (b) Profile obtained from AFM (d) Micro-Raman spectrum of the RGO reduced *in situ* by hydrazine.

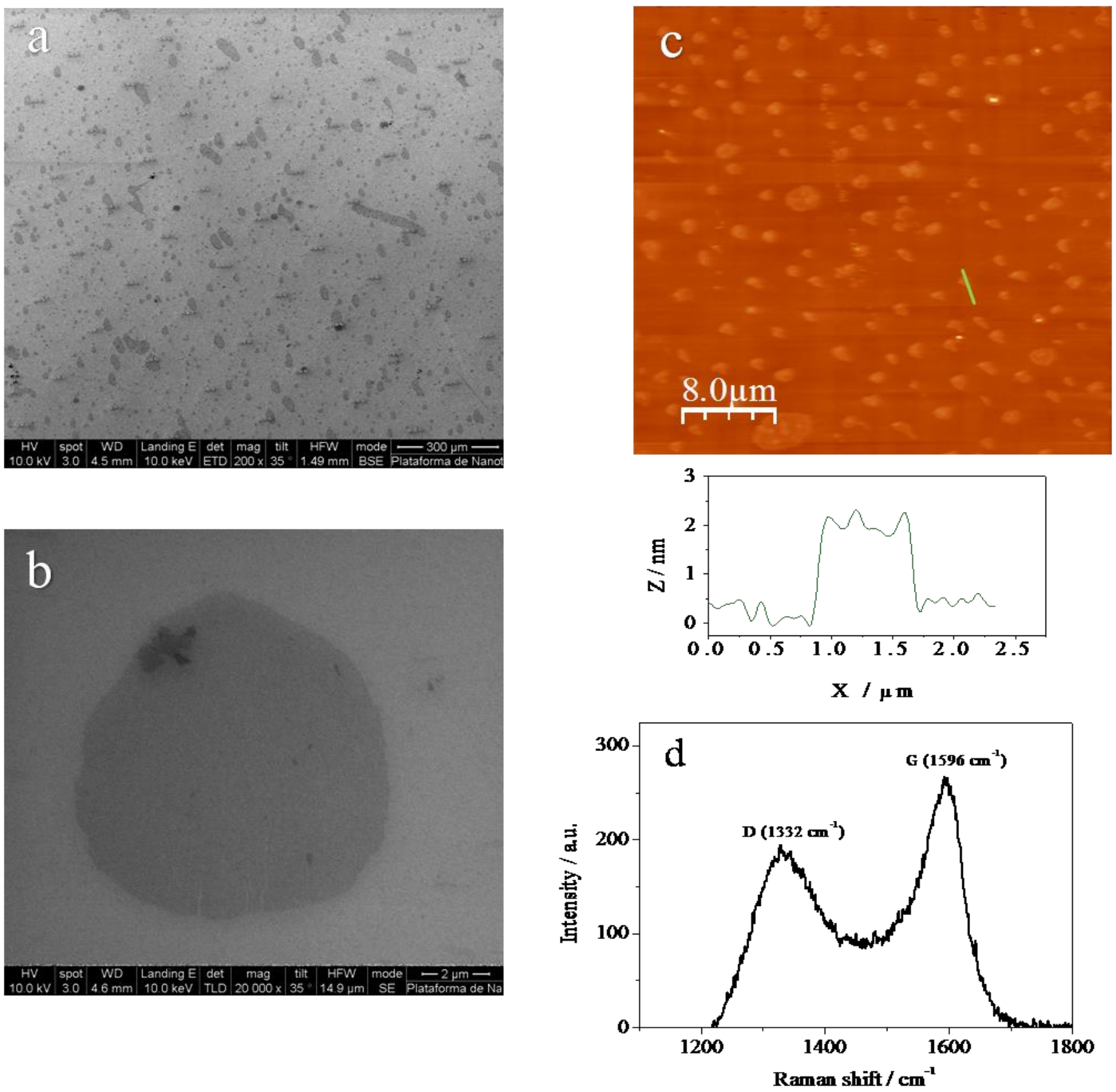

**Figure 4.** FE-SEM (a, b) and AFM (c) images of the Langmuir-Blodgett films of RGO reduced with Vitamin C. Below the AFM image is the profile obtained from AFM. (d) Micro-Raman spectrum of a representative RGO flake.

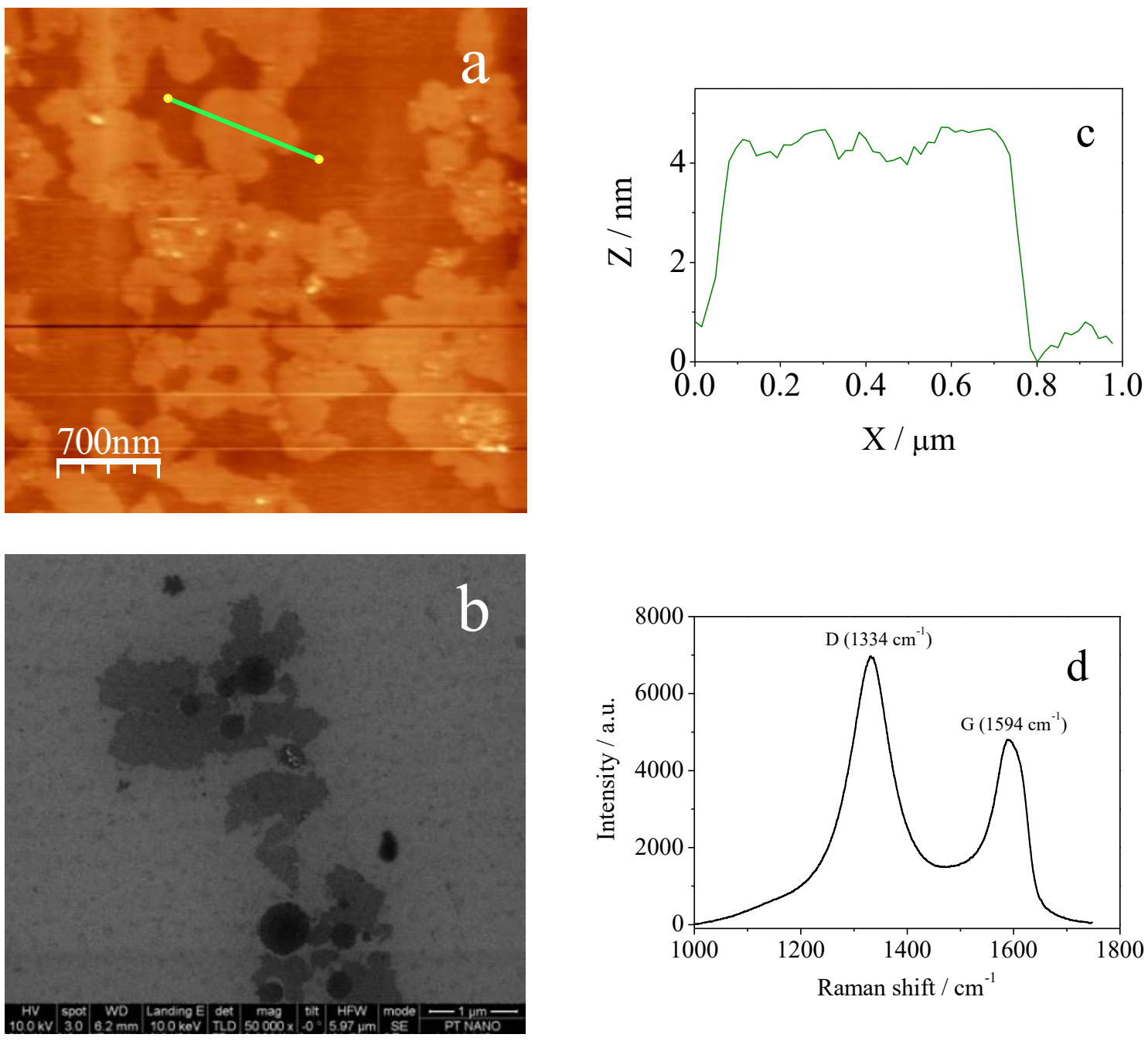

**Figure 5.** AFM (a) and FE-SEM (b) images of the Langmuir-Blodgett films of RGO reduced by hydrazine dissolved in DDPS surfactant solution. (c) Profile obtained from AFM and (d) Micro-Raman spectrum of a representative RGO flake.

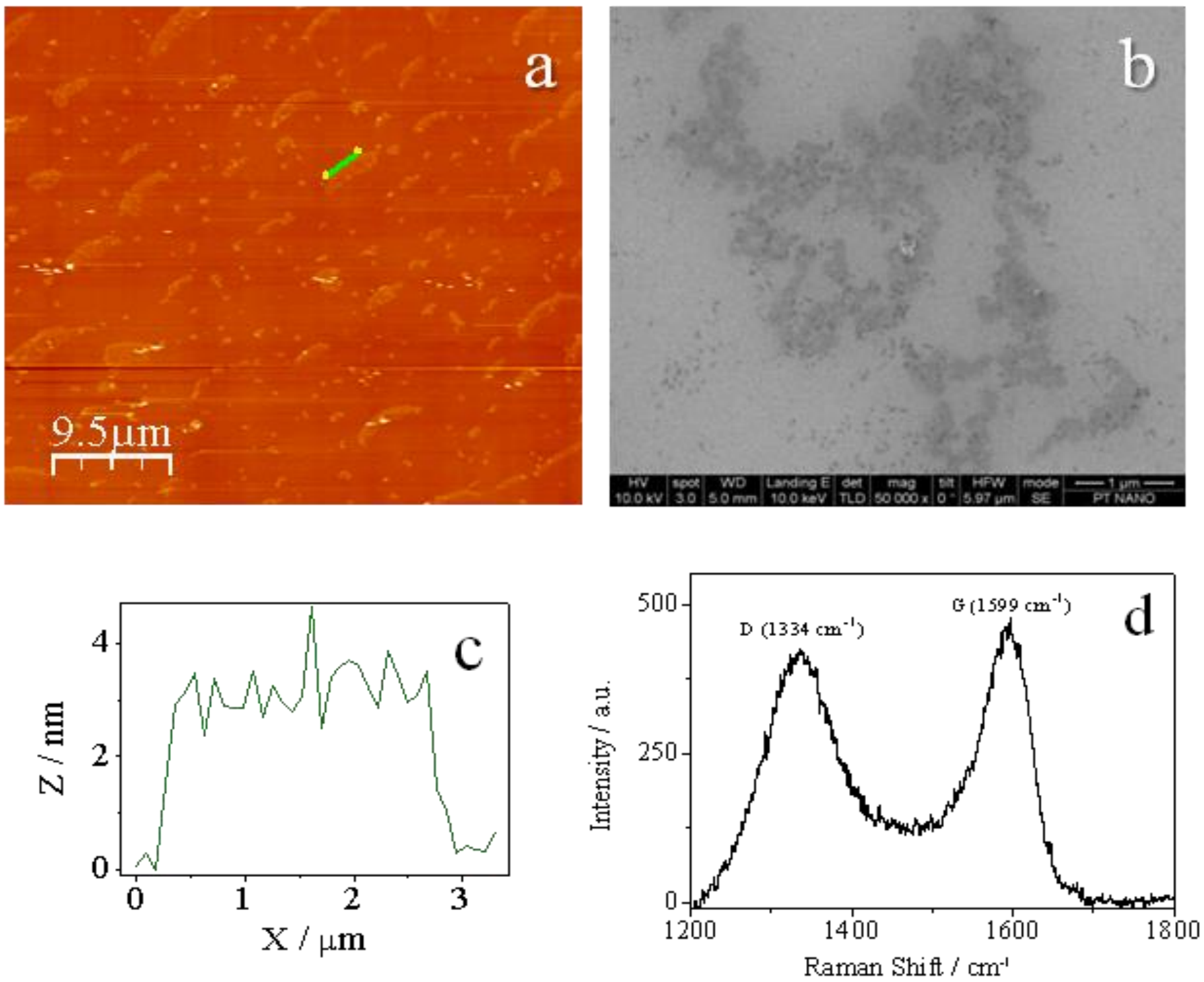

**Figure 6.** AFM (a) and FE-SEM (b) images of the Langmuir-Blodgett films of RGO reduced by Vitamin C dissolved in DDPS surfactant solution. (c) Profile obtained from AFM and (d) Micro-Raman spectrum of a representative RGO flake.